\documentclass[%
 reprint,
 amsmath,amssymb,
 aps,nofootinbib,   
]{revtex4-1}

\usepackage{bm}
\PassOptionsToPackage{linktocpage}{hyperref}
\usepackage[hyperindex,breaklinks]{hyperref}
\usepackage{enumitem}
\usepackage{slashed}

\usepackage{xcolor}
\usepackage{float}

\usepackage{array}
\usepackage{mathtools}

\usepackage{etoolbox}

\begin{document} 

\title{Gravity as universal UV completion: Towards a unified view of Swampland conjectures}
 
\author{ C\'esar G\'omez} 
\affiliation{Instituto de F\'{i}sica Te\'orica UAM-CSIC, Universidad Aut\'onoma de Madrid, Cantoblanco, 28049 Madrid, Spain}
\affiliation{
	Arnold Sommerfeld Center, Ludwig-Maximilians-Universit\"at, Theresienstra{\ss}e 37, 80333 M\"unchen, Germany
}
\affiliation{
	Max-Planck-Institut f\"ur Physik, F\"ohringer Ring 6, 80805 M\"unchen, Germany
}

\begin{abstract}
On the basis of the idea that gravity defines a universal UV-completion and using a form of UV/IR correspondence we revisit some Swampland conjectures, in particular de Sitter and infinite distance conjectures, from the point of view of renormalization group energy flow. The rol and limitations of  non perturbative instanton effects to define a metastable cosmological constant are briefly discussed in this context.

\end{abstract}
\maketitle

\section{Introduction}
 The goal of this brief note is to stress the view that quantum gravity should be thought as a universal UV completion. This point of view was put forward in \cite{us1,us2,us3,us4,us5,us6} and is based on the idea of self completion of gravity through the dynamics of classicalization. We will show that this point of view sheds some unifying light on some Swampland conjectures \cite{swan1,swan2,swan3,swan4,swan5,swan6,swan7,swan8} (for a complete list of references see \cite{palti}). These conjectures intend to identify what constraints, a consistent coupling to gravity, imposes on the low energy physics. In this note we shall suggest that for any well defined quantum field theory gravity itself defines its UV completion i.e.~it determines its infinite energy limit.

\section{An heuristic argument about UV completion and string theory}
Let us work in weakly coupled string theory with fixed $g_s^2$  smaller than one. Now consider string states lying in a Regge trajectory with masses 
given, in the limit of high level $n$, by
\begin{equation}
M^2(n) =\frac{n}{L_s^2}\text{ ,}
\end{equation}
for $L_s$ the string length. The gravitational size of one of these states is given by
\begin{equation}
R_{gr}(n) = \sqrt{n} g_s^2 L_s\text{ ,}
\end{equation}
where we use
\begin{equation}
g_s^2 = \frac{L_P^2}{L_s^2}\text{ .}
\end{equation}
Let us now introduce an effective coupling $\alpha$ by
\begin{equation}
\alpha \equiv \sqrt{n} g_s^2\text{ ,}
\end{equation}
which is the analog of t'Hooft coupling in Yang Mills  but now using the level $n$ instead of the number of colors.

The states in the Regge trajectory with $\alpha \leq 1$ correspond to string states of typical size $L_s$. However for $n$ such that $\alpha >1$ we have black holes. 

Let us first consider the string states with $\sqrt{n} g_s^2 \leq 1$. All these states have an entropy that in the limit of large $n$ {\it saturates} Bekenstein bound \cite{Bek}
\begin{equation}
S_B= \frac{M(n) L_s}{\hbar}= \sqrt{n}\text{ .}
\end{equation}
Indeed the string state at level $n$ has an entropy $S_{str}(n)$ defined by the log of the number of partitions of $n$ that, in the large $n$ limit goes qualitatively as $\sqrt{n}$, as a consequence of Hardy Ramanujan formula.\footnote{For a recent discussion on the potential connection between Bekenstein bound and unitarity in the context of quantum field theory see \cite{Gia1}.}

Thus, if we define $n_s$ as
\begin{equation}
g_s^2 =\frac{1}{\sqrt{n_s}}\text{ ,}
\end{equation}
then all states with $n\leq n_s$ and $n$ large saturate the Bekenstein bound {\it although they are not black holes}. When we reach $n=n_s$ or equivalently $\alpha=1$ the string state becomes a black hole (this is the well known string black hole correspondence point \cite{Pol}) and the Bekenstein entropy becomes equal to the Bekenstein Hawking entropy 
\begin{equation}
S_{BH} = \frac{L_s^2}{L_P^2}\text{ .}
\end{equation}

When we enter into the black hole regime
$\alpha >1$ what we encounter is that the combinatorial string entropy $S_{str}(n)$  that goes like $\sqrt{n}$ is smaller than the Bekenstein Hawking entropy 
\begin{equation}
S_{BH}(n) = \alpha \sqrt{n} = \alpha S_{str}(n)\text{ .}
\end{equation}
This means that in the {\it strong coupling regime} defined by $\alpha >1$ the string degrees of freedom are, at least superficially, unable to account for the black hole  entropy. 

In other words, the high energy regime $\alpha >1$ is described by pure gravity that in this sense UV completes the weakly coupled string theory. In essence what this means is that for energies larger than $\frac{M_s}{g_s^2}$ i.e.~for $\alpha >1$ the string theory needs to be unitarized by purely gravitational degrees of freedom. 

The simplest way to avoid the former conclusion and to maintain, from this point of view, string theory as UV complete requires to introduce a {\it running string coupling} such that at arbitrary high energies $g^2_s(E)$ flows into {\it the fixed point condition} $\alpha=1$. This condition implies
\begin{equation}\label{one}
g_s^2(E) = \frac{1}{\sqrt{n(E)}}\text{ ,}
\end{equation}
with $\sqrt{n(E)} = E L_s$. Let us now translate this relation in terms of the dilaton potential $V(\phi)$. In order to do it we define
\begin{equation}
N(\phi) \equiv \frac{M_P^4}{V(\phi)}\text{ .}
\end{equation}
This number is simply the Gibbons Hawking entropy for a cosmological constant $\Lambda(\phi) = \frac{V(\phi)}{M_P^2}$. Identifying $N(\phi) = \sqrt{n(E)}$ and using the fixed point condition (\ref{one}) we get
\begin{equation}
g_s= \sqrt{V} L_P^2\text{ .}
\end{equation}
That using now the dilaton representation $g_s =e^{-\phi}$ yields
\begin{equation}
V(\phi) \sim M_P^4 e^{-2\phi}\text{ ,}
\end{equation}
that leads to the so called de Sitter conjecture \cite{swan5} $|V'| \sim V$. The identification $N(\phi)=\sqrt{n(E)}$ associates with the energy level $n(E)$ the formal cosmological constant $\Lambda(\phi)$ whose Gibbons Hawking entropy is the same as the string entropy of the corresponding Regge state at level $n(E)$. This {\it black hole-cosmological constant correspondence} is a UV/IR correspondence that associates with the UV high energy black hole in the Regge trajectory an IR cosmological constant with Hubble radius given by the corresponding black hole radius.  Hence in this simple approach the constraint on the dilaton potential $|V'| \sim V$ appears as the IR version of the UV {\it self unitarization of string theory at high energies}. In the next paragraph we shall consider other Swampland conjectures.

\section{Some comments on Swampland conjectures}
If we decide that a quantum field theory is consistent  only if the deep UV is controlled by pure gravity we get a new view on some Swampland conjectures.

\subsection{Weak gravity conjecture}
Let us consider first the {\it weak gravity conjecture} introduced in \cite{swan2}. Our argument will be a slightly modified version of the one in \cite{swan2}. 
The existence of a force weaker than gravity implies the existence of an energy scale $M_{W}$ larger than $M_P$. This scale is defined by the distance at which the weakest force becomes strongly coupled and is given by $\frac{M_P}{e}$ for $e$ the coupling defining the weakest force. Associated with this energy scale we can define $N(W) =\frac{r(W)^2}{L_P^2}$ with $r(W) =M(W) L_P^2 =\frac{L_P}{e}$ (we are using the convention $\hbar =1$). Note that $M(W)$ can be interpreted as the mass of an extremal black hole. 

In \cite{swan2} it was assumed that this extremal black hole, in absence of SUSY, should have channels of decay not determined by Hawking evaporation. This assumption forces the existence of an elementary charged particle satisfying $m\leq eM_P$. We will instead address the problem from the point of view of gravity self completion. Self completion  implies the existence of an elementary charged particle defining the {\it gravitational constituency} of the black hole of mass $M_W$. This sets the mass of this particle to satisfy $mN(W) \leq M(W)$ which is the relation in \cite{swan2}. The potential decay assumed in \cite{swan2} has the natural interpretation of quantum depletion introduced in \cite{us3}.

\subsection{Distance conjecture}

Let us now consider the {\it infinite distance conjecture} \cite{swan3}. Our approach will consist in defining the distance using {\it renormalization group flow in energy}. Using the relations of the former paragraph we define for any energy scale $E$
\begin{equation}
N(E) \equiv \frac{R_{gr}(E)^2}{L_P^2}\text{ .} 
\end{equation}
Let us now define a {\it distance between energy scales} as
\begin{equation}
d(E',E) = \ln \left(\frac{N(E')}{N(E)}\right)\text{ .}
\end{equation}
Using these definitions we observe that infinite energy is at infinite distance, namely $N(\infty)=\infty$. Using the black hole-cosmological constant correspondence we can associate with this distance in energy a distance in field space. Namely from $N(E)= e^{2\phi}$ we get $d(E',E) = 2 (\phi(E')-\phi(E))$ for $N(\phi(E)) \equiv \frac{M_P^4}{V(\phi)}$.

Let us now define  the mass scale $m(E)$ by the relation $m(E)N(E) =E$ that yields in terms of the dilation field
\begin{equation}
m(\phi) = M_P e^{-\phi}\text{ .}
\end{equation}
Now we can compare the former expression with the distance conjecture relation \cite{swan7}
\begin{equation}
m=|\Lambda|^{\alpha} = M_P e^{-2 \alpha\phi}\text{ ,}
\end{equation}
with $M_P^2\Lambda= V$. This leads to $\alpha=1/2$. In the infinite energy limit, if the theory is UV completed by gravity, we get $N(\infty)=\infty$ and the corresponding tower of massless states.

\subsection{de Sitter conjecture and non perturbative KKLT effects}

Finally let us consider the {\it de Sitter conjecture} \cite{swan3}. The former argument implies that any classical background geometry defining a vacuum should solve the fixed point condition
\begin{equation}
\frac{dN}{dE}{\bigg|}_{E_c} =0\text{ ,}
\end{equation}
for some energy $E_c$. Using the representation in terms of $\phi$ this relation becomes $\frac{dN(\phi)}{d\phi}=0$. The existence of a classical cosmological constant requires that this fixed point happens for some finite $E_c$. If this is the case we can have two options for this fixed point. Either $N(E_c)=\infty$ meaning that this classical background is at {\it infinite distance} or $N(E_c)$ is finite.

Let us consider now both cases separately. In the second case i.e.~when the fixed point is at {\it finite distance} the fixed point represents a full fledged quantum state with finite $N$ that will departure from classicality in a finite quantum breaking time. This has been discussed in \cite{us6,cc1} and represents the {\it quantum breaking time de Sitter conjecture}.\footnote{For some phenomenological consequences of quantum breaking see \cite{Zell1}.} 

The second possibility namely $N(E_c)=\infty$ corresponding to a {\it classical background} with $V'=0$ and $V$ finite
comes with a tower of massless states of mass $m\sim \frac{V}{N}$. This tower of states reduces the gravitational cutoff \cite{Gia} to $\frac{M_P}{\sqrt{N}}$. Thus if we insist in keeping the corresponding length scale $\sqrt{N}L_P$ at least of the order of  $L_s$ we need to have $g_s^2 N = O(1)$.

This last comment is important to identify the limits of some attempts to find a classical background with a positive cosmological constant. In this case the potential $V(\phi)$ should be modified adding some non perturbative effects. If these are of instanton type, as it is the case in KKLT like models \cite{KKLT}, their {\it four dimensional} contribution $O(e^{-\frac{1}{g_s}})$ will be suppressed as $e^{-\sqrt{N}}$. 

In other words, a classical de Sitter space time with $V'=0$, $V$ finite but $N=\infty$ cannot be obtained using standard instanton non perturbative effects if the effective gravity cutoff scale is of order $L_s$ or smaller. In other words, we can conjecture that {\it non perturbative instanton effects contributing to the stability of a classical background are suppressed  as} 

\begin{equation}
e^{-\sqrt{N}}\text{ ,}
\end{equation}
{\it with $N\sim e^{d}$ for $d$ the distance, in the former renormalization group sense, at which the classical background is located}. If this distance is finite the corresponding background has intrinsic quantum instabilities i.e.~quantum breaking and if it is at infinite distance needs some non perturbative effects not suppressed in the $N=\infty$ limit.\footnote{This suppression of instanton effects is very similar to the one in Yang Mills with large number of colors. In reality if we consider instantons with fractional topological charge \cite{CG} the relevant suppression factor will go as
\begin{equation}
e^{-\frac{\sqrt{N}}{N_c}}\text{ ,}
\end{equation}
for $N_c$ the number of colors. This opens the possibility of non perturbative contributions even at infinite distance if we work in the limit 
$N_c \sim \sqrt{N}$.}

A potential fixed point solution can however take place in the case we have SUSY where the condition setting the classical background $\frac{dN}{dE}=0$ is determined by a CFT vanishing beta function. This is what probably happens in the case of $AdS_5 \times S^5$ with maximal SUSY where $N$ has the meaning of the central extension of the dual CFT and where the fixed point condition is determined by the underlying non renormalization theorems of SUSY.

In summary, we conclude that most of the Swampland conjectures can be understood using two basic assumptions. First that the UV completion of any consistent quantum field theory is defined by quantum gravity and secondly that gravity itself is self complete. As an output we have observed that classical de Sitter backgrounds at infinite distance cannot be stabilized by standard instanton effects and that those at finite distance are quantum mechanically unstable through quantum breaking. It seems that Nature pushes us into asymptotic flatness.

{\bf Acknowledgements.}
I would like to thank Gia Dvali, Dieter Lust, Eran Palti and Sebastian Zell for comments and enlightening and relevant discussions. This work was supported  by the ERC Advanced Grant 339169 "Selfcompletion" and  the grants SEV-2016-0597, FPA2015-65480-P and PGC2018-095976-B-C21. 

\end{document}